\documentclass[a4paper,fleqn,usenatbib,useAMS]{mnras}
\usepackage{graphicx}	% Including figure files
\usepackage{amsmath}	% Advanced maths commands
\usepackage{amssymb}	% Extra maths symbols
\usepackage{multicol}        % Multi-column entries in tables
\usepackage{bm}		% Bold maths symbols, including upright Greek
\usepackage{pdflscape}	% Landscape pages

\title[Cosmic Initial Condition for a Habitable universe]{Cosmic Initial Condition for a Habitable universe}
\author[Sohrab Rahvar ]{Sohrab Rahvar$^{ }$\thanks{E-mail: rahvar@sharif.edu}\\
$^{ }$Physics Department, Sharif University, P.O.Box 11365-9161, Azadi Avenue, Tehran, Iran}

\begin{document}
\label{firstpage}

\maketitle

\begin{abstract}
Within the framework of an eternal inflationary scenario, a natural question regarding the production of eternal bubbles is the essential condition requires to have a universe capable of generating life. In either an open or a closed universe, we find an anthropic lower bound on the amount of e-folding in the order of $60$ for the inflationary epoch, which results in the formation of large-scale structures in both linear and non-linear regimes. We extend the question of the initial condition of the universe to the sufficient condition in which we have enough initial dark matter and baryonic matter asymmetry in the early universe for the formation of galactic halos, stars, planets and consequently life. 
We show that the probability of a habitable universe is proportional to the asymmetry of dark matter and baryonic matter, while the cosmic budget of baryonic matter is limited by the astrophysical constrains.
 \end{abstract}

%\pacs{98.80.-k, 98.65.-r, 11.30.Fs}% PACS, the Physics and Astronomy


\section{Introduction}
%\maketitle

 The habitability of the Universe is one of the fundamental issues of cosmology. In other words, what are the specifications required for a universe to be habitable ? In terms of fundamental physics, we can think about possible different physical parameters for which our universe is adapted. This possibility might be realized in multiverse models where the physical parameters vary in an ensemble of parallel universes \citep{mult}. Here in this work, we adapt the "weak anthropic" principle, by means of that physical constants are given and we investigate the initial conditions of different universes. We study the formation of a habitable universe based on the initial conditions in the early epoch of the universe, during the inflationary period.

 In the standard inflationary model for an early universe,  a scalar field, so-called inflaton field, drives a rapid phase of expansion of the universe. This inflationary phase stretches any local features in the curvature of the early universe into spatially flat space \citep{star,guth}. Also during this inflationary phase, all primordial defects from the phase transition manifesting as topological defects are diluted \citep{infbook}. The other consequence of inflation is that all parts of the universe that were thermalized within the horizon before the beginning of inflation stretch out and makes universe uniform at super-horizon scales. This uniformity has been observed with the level of $10^{-5}$ on the map of Cosmic Microwave Background (CMB) radiation \citep{flat}.

 The other advantage of inflationary cosmology is that quantum fluctuations of the inflaton field results in seeds for the formation of large-scale structures. These small density fluctuations in the dark matter fluid, of order $\sim 10^{-5}$, eventually grow and produce potential wells and, through gravitational condensation and cooling, 
 the baryonic matter of cosmic fluid forms galaxies, stars and planets. Recent observations of CMB by the {\it Wilkinson Microwave Anisotropy Probe} (WMAP) and {\it Planck} satellites show the compatibility of the observational data with the prediction of the inflationary models \citep{planck,wmap}.
 
 While inflationary cosmology is successful in its explanation of flatness, homogeneity and power spectrum of large scale structures, there are challenging questions for this theory, such as the entropy problem. Since the entropy of 
 all systems as well as the universe increase with time, the early universe must start with very low entropy \citep{penrose}, which means that fine tuning for the initial conditions of the universe might be needed. A possible solution to this problem is the chaotic inflationary model introduced by \citet{linde}. In this model, the initial condition of the universe from a large pre-inflationary domain has arbitrary uncorrorlated values within Planck-size patches and a successful universe with low entropy at a given patch produces a sufficient large domain. The parameter associated with this expansion is defined 
by the e-folding number, which is the logarithm of the ratio of the scale factor at the end to that at the beginning of the inflation ( i.e. $N = \ln{a_f}/{a_i}$). For a large value of e-folding, the spatial part of curvature turn to be nearly flat. For each inflationary area, the quantum fluctuation of the inflaton field can be larger than the classical decline of the field. The result would be to produce new inflationary areas, or in other words bubbles,  out of a parent domain \citep{etenral}. The consequence of eternal inflation is the production of infinite inflationary areas, of which some can meet the condition for the formation of life.

For an unsuccessful inflationary phase owing to an insufficient number of e-folding, the density of the universe after the end of inflation, depending on the initial conditions, could be either dense or dilute compared with a flat universe. As a result, after inflation ends, the universe would either collapse in a short cosmological time-scale or expand and dilute, with no chance for the formation of structures. After the end of inflation, all the energy in the inflaton field turns into the matter and radiation during the reheating \citep{reheating}. For a sufficient asymmetry between baryons and anti-baryons, the remaining baryonic matter after annihilation can form the baryonic structures within the dark halos.

 In this work, we investigate the inflationary scenario and post inflation epoch from the point of view of the anthropic principle. This kind of investigation has been started in the early days of developing modern cosmology. \citet{dicke} noted that the age of the Universe cannot be a random value: biological factors constrain it to be not too young and not too old. A young universe does not have enough metals for the formation of life and in an old universe all the stars would have left the main sequence. The effect of the cosmological constant on the formation of structures within the framework of anthropic principle has also been studied by \cite{winberg}. He concluded that having a larger value for the cosmological constant prevents the collapse of structures and formation of galaxies and stars. A general study of the set of physical constants 
 and parameters that support life can be found in \cite{tegmark1}, as well as detailed discussion of the anthropic principle in \cite{b1}.

 In the first part of this work, we study the prior probability of the formation of structures from the initial condition of the inflationary scenario.  We then discuss the selection condition to have enough baryonic number in the universe for the formation of galaxies, stars, planets and consequently life. In Section (\ref{inf}), we briefly introduce inflationary cosmology, with an emphasis on eternal inflation.  Here we investigate the initial condition for the pre-inflationary patches to form a habitable universe. In Section (\ref{structure}), we investigate the formation of dark matter structures in open and closed universe, taking into account e-folding from inflation and matter asymmetry of the universe. The failure of dark matter structure formation means no formation of stars and planets. In Section (\ref{ba}), we study the baryon asymmetry of the universe as sufficient condition for the formation of structures, stars and planets within the gravitational potential of dark matter. A conclusion is given in Section (\ref{conc}).
 
 \section{chaotic inflationary cosmology and probability of formation of a universe with non-zero curvature}
 \label{inf}
 
The flatness and horizon problems of the Universe were important issues in the standard big bang cosmology at the beginning of the 1980's. As a solution, in the inflationary model, a rapid expansion of the early phase after the big bang could resolve these problems. In the old picture of the inflationary scenario, a phase transition could provide energy for the rapid expansion of the Universe, while the energy density of the Universe for this phase remains constant \citep{guth}. However, this scenario had the problem of bubble collision if inflationary areas were close to each other; they were far from each other if a smaller value for $\Omega$ was considered. In the new scenario, instead of phase transition in a scalar field, the scalar field undergoes a slow-rolling dynamics and can resolve the above mentioned problem. Moreover, quantum fluctuations of the scalar field produce density fluctuations that are statistically compatible with observations \citep{linde82}.

As a brief introduction to the inflation dynamics, let us assume a scalar field that drives the dynamics of the early universe. For a scalar field with Lagrangian ${\cal L} = 1/2\partial_\mu\phi\partial^\mu\phi - V(\phi)$, we assume the slow-rolling condition, where the kinetic energy is negligible compared with the potential term. Also, we ignore the spatial components of the Lagrangian, as they are diluted by fast expansion of the universe. This condition can also be given by two parameters, so-called slow-rolling conditions of  $\epsilon = m_{pl}^2(V^{'}(\phi)/V(\phi))^2/2\ll 1$ and $\eta = m_{pl}^2|V^{''}(\phi)/V(\phi)|\ll 1$, which implied the  Friedmann-Robertson-Walker 
 (FRW) equation; the continuity equation simplifies to 
 $H^2 \simeq 8\pi V(\phi)/3m_p^2$ and $3H\dot{\phi} = - V^{'}(\phi)$. Here, for simplicity we take natural units.

 One of the main characteristics of inflationary models is the number of e-folding and its dependence on the initial conditions of the scalar field at the beginning of the inflationary phase of the universe. Within the framework of chaotic inflation, different domains of the universe at the Planck time have the scalar fields with stochastic distribution and, for a sufficiently large scalar field that satisfies the slow-rolling condition, inflation can start. If we treat $\phi$ as the classical scalar field, starting from an arbitrary position in phase space (i.e. ($\phi,\dot{\phi}$)), the well-known classical attractor tends asymptotically the scalar field to hold the slow-rolling condition, whereas for quadratic and quartic potentials there is a lower bound on the initial value of scale field to produce successful inflation \citep{Yi93}. 

Here, the $\dot\phi$ term, which is related to the kinetic energy, is also produced by the quantum fluctuations. 
If the quantum fluctuations of inflatons become larger than the potential term, then we can ignore the potential term in the density and pressure of the inflaton field and the equation of state would be $p\simeq \rho$. Then, from the continuity equation, the energy density or the kinetic energy term of the inflaton field decreases as $1/2\dot\phi^2 \propto 1/a^6$, much faster than the radiation. The result is that the slow-rolling condition can hold for a shorter time-scale. We can also consider quantum fluctuations of the inflaton field as the stochastic motion of the field in phase space. This effect shrinks the proper domain of the phase space for inflation. We note that a classical slow-rolling attractor in  phase space is almost immune to stochastic quantum fluctuations of the inflaton field and the initial conditions of the inflaton field are almost independent of $\dot{\phi}$ \citep{grain}.
 
Important observational constrains for the scalar field can be imposed by observation of CMB radiation. The {\it Planck} satellite could constrain various 
models of inflationary potentials. Here we adapt a power-law potential for our study  \citep{linde}, 
\begin{equation}
V(\phi) = \lambda m_{pl}^4 (\frac{\phi}{m_{pl}})^n,
\end{equation}
where CMB observations constrain the power-law index to be in the range of $n<2$ \citep{planck}. Our aim is to investigate conditions in this model to have a habitable universe. Let us assume a uniform probability for the energy of inflaton field \citep{Yi93}, with energy smaller than the Planck scale (i.e.  $\rho_\phi<m_{pl}^4$).
The constrain on energy density applied to the initial conditions imposes that scalar field should be in the range of $0<\phi_i<m_{pl}/\lambda^{1/n}$. Then the prior probability of a scalar field at the beginning of inflation is obtained by 
 \begin{equation}
\frac{dP_{prior}(\phi)}{d\phi} = \frac{n\lambda}{m_{pl}}(\frac{\phi}{m_{pl}})^{n-1}.
\label{prob}
\end{equation}
For a generic potential the number of e-folding \citep{infbook} is calculated  by
\begin{equation}
N \simeq \int^{\phi_i}_{\phi_{end}}\frac{V}{V'} \frac{d\phi}{m_{pl}^2}, 
\end{equation}
where, for the power-law potential, the result of the integration is 
\begin{equation}
\nonumber
N = \frac{4\pi}{n m_{pl}^2}(\phi_i^2 - \phi_{end}^2)
\end{equation}
 \citep{mukhanov} and, by defining the end of inflation at  
$|V^{''}(\phi)|\ll 9H^2$
\citep{stein}, the corresponding field for the end of inflation is 
\begin{equation}
\nonumber
\phi_{end} = \sqrt{|\frac{n(n-1)}{24\pi}|} m_{pl}.
\end{equation}
 This means that the initial condition of the inflation in terms of number of e-folding is given by 
\begin{equation}
\phi_i = \sqrt{n}m_{pl}(\frac{n-1}{24\pi} + \frac{N}{4\pi})^{1/2}\simeq \sqrt{\frac{Nn}{4\pi}} m_{pl},
\end{equation}
 where for a large $N$ we can ignore the term corresponds to the end of inflation. 
Let us assume a minimum value of e-folding is required for the formation of life, we call it $N_\star$. Then the initial condition for the inflaton field must satisfy the condition of $\phi_i \geq \sqrt{N_\star n/4\pi} m_{pl}$.
The assumption that the energy associated with this field is smaller than the Planck 
energy imposes an upper bound on the initial value, $\phi_i<m_{pl} \lambda^{-1/n}$. Then the 
habitable initial condition of a scalar field should be within the following range:  
\begin{equation}
\sqrt{\frac{N_\star n}{4\pi}}<\phi_i/m_{pl}< \lambda^{-1/n}, 
\end{equation}
where, integrating from equation  (\ref{prob}), the probability of scalar field to be larger than $N_\star$ is
 \begin{equation}
 P_{prior}(>N_\star) = 1-\lambda(N_\star n/4\pi)^{n/2},
 \label{pp}
 \end{equation}
 where the  maximum number of e-folding by the condition of $P_{prior}=0$  is $N_{max} = 4\pi \lambda^{-2/n}/n$. This means that, for a habitable universe, e-folding 
must be in the range $N_\star<N<N_{max}$. 
%We note that the probability of having a habitable universe decreases by increasing the e-folding number. 

From the Planck data, the indices compatible with observations are $n=4/3, 1$ and $2/3$ \citep{planck}.  We can also identify the coupling constant of scalar field (i.e. $\lambda$) from 
the temperature fluctuations in the CMB. The quantum fluctuation of the scalar field produces fluctuations in the cosmic fluid, with amplitude \citep{ll}
\begin{equation}
A_s^2 = \frac{512 \pi}{75 m_{pl}^6} \left.\frac{V^3}{V'^2}\right\vert_{k=aH}
\end{equation}
where, for the power-law potential, the amplitude is 
\begin{equation}
A_s^2 = \frac{512 \pi}{75} \left.\frac{\lambda}{n^2} (\frac{\phi^{n+2}}{m_{pl}^{n+6}}) \right\vert_{k=aH},
\end{equation}
and using the value of $A_s$ from CMB observations and substituting the scalar field for the beginning of inflationary period results in 
\begin{equation}
\lambda=\frac{75n^2}{512 \pi}A_s^2\left(\frac{nN_{60}}{4\pi}\right)^{-\frac{n+2}{2}},
\label{lambda}
\end{equation}
where, from the recent observations by Planck satellite, the amplitude of perturbation is $\ln(A_s^2 10^{10}) = 3.094$ and the number of e-folding to generate the present map of the CMB is $N_{60} \sim 60$. Using $A_s$ in equation (\ref{lambda}) results in the numerical value of $\lambda$ as one of the natural constants. Substituting equation (\ref{lambda}) in the probability function (\ref{pp}), the probability of a habitable universe in terms of $N_\star$ and $n$ is given by 
\begin{equation}
P_{prior}(>N_\star) \simeq 1 - \frac{5n}{512}(\frac{N_\star}{60})^{n/2}A_s^2.
\label{e}
\end{equation}
For the case of e-fold in the order of $N\simeq 60$, which provides an age for the universe 
almost comparable to ours, the probability from equation (\ref{e}) is almost unity, while for a universe with a very large $N_\star$ the probability tends to zero. 
 \begin{figure}
\includegraphics[width=85mm]{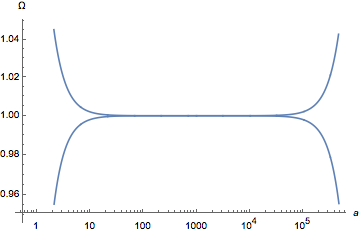}
\caption{Demonstration of the variation of $\Omega$  with the initial condition of $\Omega_i<1$ and 
$\Omega_i>1$ from the beginning to the end of inflation and the deviation of $\Omega$ from unity after the end of inflation. $x$-axis represents the scale factor, $a$, and the $y$-axis 
represent $\Omega$. Here we take the number of e-folding smaller than the conventional value for better demonstration of variation of $\Omega$ with the scale factor.}
\label{fig1}
\end{figure}

The consequence of smaller e-folding is that the universe cannot approach a flat space and a direct effect of this is on the formation of large-scale structures of the universe.
Let us assume a patch of the universe before the beginning of inflation with initial curvature $\Omega_i \neq1$.
The number of e-folding before reheating is $\ln(a_f/a_i) = \frac{4\pi}{n}(\phi_i^2 - \phi_f^2)$ \citep{mukhanov}. 
Since the spatial comoving curvature of the universe (i.e. $k$) is time-independent, the final value of $\Omega$ at the end inflation is
 \begin{equation}
 \Omega_f = 1 + (\frac{a_i}{a_f})^2(\Omega_i - 1),
 \end{equation}
 where, during the slow-rolling of the scalar field, we take $H$ almost constant. On the other hand, after inflation, assuming a radiation-dominate epoch for the universe, $\Omega$ of the universe changes with the scale factor as 
 \begin{equation}
 \Omega(a) = 1 + (\frac{a}{a_f})^2(\Omega_f - 1).
 \end{equation}
Fig. (\ref{fig1}) is a representation of the evolution of $\Omega$ as a function of the scale factor from the beginning to the end of inflation and after the inflationary epoch. Having smaller e-folding, depending on the sign of the curvature, causes a universe to dilute or collapse on a shorter time-scale, with not enough time for the formation of structures. 

In the next section, we examine the condition for the initial condition of 
inflation for open and closed universes that lead to the formation of  large-scale structures. We note that a consequence of the structure formation in the linear regime is the formation of non-linear structures and eventually the formation of stars, planets and life.

\section{Structure formation and constrains on initial condition}
In this section we study constrains on the initial condition 
of the universe for the formation of structures in open and closed universes.

\label{structure}

 \subsection{Closed universe}
 In a closed universe, any overdensity with respect to the background (i.e. $\delta_i>0$) due to the negative energy condition will grow with time. To prove this argument,  we take the total energy of an overdensity region as 
\begin{equation}
E = K_i -K_i\Omega_i(1+\delta_i),
\end{equation}
where the first term $K_i$ represents the initial kinetic energy and the second term is the potential energy. The energy condition of $E<0$, imposes the criterion 
$\delta_i > \Omega_i^{-1} - 1$, which allows for collapse of a structure. In the case of $\Omega_i\geq 1$ and $\delta_i>0$ the collapse condition is always satisfied. A key parameter for the habitability of the universe in this case is the age of the universe. In a  high-aged universe, structures have opportunity to turn from the linear to the non-linear phase and finally stars and planets can form.

Let us assume a closed universe with $\Omega_i>1$ at the pre-inflation phase. The universe will turn around at a given scale factor with the condition $H(a_t) = 0$, where the subscript $"t"$ is assigned 
to turnaround time.  Then FRW 
equation at this point simplifies to $k/a_t^2 = 8\pi\rho(a_t)/3$, where the total density at the time of turnaround is related to the density of matter and radiation at the end of inflation (i.e. $\rho_f$) as $\rho(a_t) = \rho_{R,f}(a_t/a_f)^{-4} + \rho_{M,f}(a_t/a_f)^{-3} $. Using the definition of $k=(\Omega_f-1)a_f^2 H_f^2$ at the end of inflation, the ratio of turnaround scale factor to the scale factor at the end of inflation is determined by  
\begin{equation}
\frac{a_t}{a_f} = \left(\frac{\Omega_{R,t} + \Omega_{M,t}}{\Omega_{R,t}(\Omega_i - 1)}\right)^{1/2} \frac{a_f}{a_i},
\label{omg}
\end{equation}
where $\Omega_{R,t}$ and $\Omega_{M,t}$ are the ratio of density of radiation and matter to the critical density of the universe at the turnaround time. Assuming Grand Unified Theory (GUT) energy for the end of inflation \citep{gut} of the order of $10^{16}$ Gev and expressing the temperature of the universe in terms of the temperature of CMB,  e-folding 
of inflation in terms of the parameters of a closed universe is calculated by 
\begin{equation}
\ln\frac{a_f}{a_i} = \ln\frac{10^{16} \text{GeV}}{T_{R,t}} + \frac12\ln\frac{\Omega_{R,t}(\Omega_i -1)}{\Omega_{R,t} + \Omega_{M,t}},
\label{efolding}
\end{equation}
where $T_{R,t}$ is the temperature of the CMB at turnaround time. We note that the turnaround temperature is the minimum temperature of a closed universe.

\begin{figure}
\centering
  \includegraphics[width=80mm]{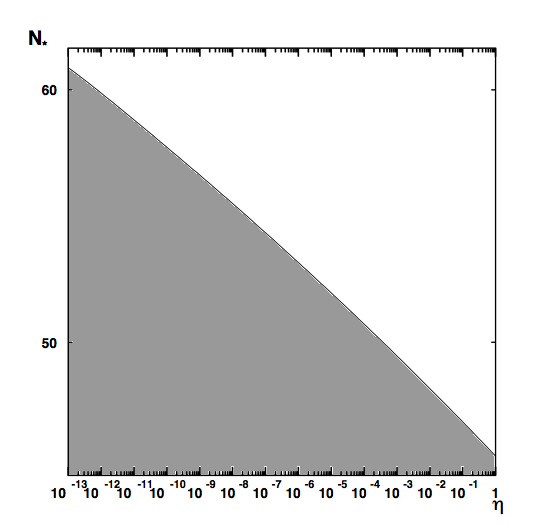}
  \caption{Exclusion area identified by grey colour in terms of $N_\star$ and $\eta$ in a closed universe where the temperature of the universe is smaller than $T_{max}$ ( the maximum temperature that allows for the formation of life). This condition results from equation (\ref{cc}).}
  \label{open}
\end{figure}

Now, we investigate a maximum temperature of the universe (i.e. $T_{max}$) that 
enables the formation of stars, planets and life. For a universe with $T<T_{max}$, the habitable condition is satisfied in general. The 
time-scale for the formation of life should be at least larger than the 
time-scale for the planet formation. In terms of star formation, this should be at least after the second generation of stars where heavy elements have already been synthesized. Moreover, at the time of formation of life, there should not be deleterious radiations in $UV$, $X$ and $\gamma$ wavelengths to dissociate organic molecules \citep{dayal}. Recently, a higher maximum temperature of the universe that allows for the formation of life has been suggested. We may assume a universe at the temperature of $T_{max} \sim 300$K with the cosmic ambient temperature comparable to that of the Earth  \citep{loeb} or possible carbon-made planets at redshift $z<50$ \citep{mashin}. The main problem with these models is the lack of existence of heavy elements as carbon, nitrogen, oxygen, phosphorus and sulphur \footnote{ The six element so-called CHNOPS are the backbone of life} \citep{astrobio} in locations where they are essential for the formation of life.

 An astrophysical constrain for $T_{max}$ would be around $z\simeq 1$ which can be translated in terms of temperature as  $T_{max}\simeq 10^{-4}$ eV, where the star formation burst epoch is near to its end \citep{loeb2}.  The extreme case of $T_{max} \simeq 300$K, as suggested by \cite{loeb},  is equivalent to $T_{max}\simeq 10^{-2}$ eV. Since in a closed universe the lowest temperature happens at the turnaround time, we set $T_{max} = T_{R,t}$.

We use the general definition $\eta = n_X/n_\gamma$, where $n_X$ is the number density of dark matter particles ,and rewrite this parameter as follows:
\begin{equation}
\eta = \frac{T}{m_X}\frac{\Omega_M}{\Omega_\gamma}.
\label{eta1}
\end{equation}
Substituting this equation in (\ref{efolding}) results in 
\begin{equation}
\ln\frac{a_f}{a_i} = \ln\frac{10^{16} \text{GeV}}{T_{max}} + \frac12\ln\frac{T_{max} (\Omega_i -1)}{ T_{max}+ m_X\eta},
\label{cc}
\end{equation}
where we assume $\Omega_i$ is of the order of unity. For the dark matter particles with the mass of $m_X \simeq 10$ GeV \citep{april} and $T_{max} \simeq 10^{-4}$ eV, we plot the e-folding in terms of $\eta$ in Figure (\ref{open}), where in the grey area, temperature is higher than $T_{max}$ and we exclude this part. From equation (\ref{e}), the probability of having a habitable universe with e-folding larger than this range is high.

 \subsection{Open universe}
 
Here, we study the possibility of formation of large-scale structures that consequently produce stars and planets in an open universe. 
The  crucial parameter in this universe is the background density of the universe compare to the expansion rate of the universe; when this is small enough, density perturbations dilute very quickly and prevent the growth of the initial perturbations after the end of inflation.

The initial perturbation of the scalar field in the inflationary area is produced by the quantum fluctuations of the scalar field. This perturbation affects the metric and energy-momentum tensor of the cosmic fluid. The wavelength associated to the quantum fluctuations during the inflationary 
epoch grows exponentially and, after exiting the horizon, due to damping term from the expansion of the universe, quantum fluctuations freeze out and behave classically. After the end of inflation, the freezing modes re-enter the horizon with a density contrast of the order $10^{-5}$ and start growing due to the gravitational instability.

Let us assume that a perturbation mode reenters to the horizon. 
In this case, from the standard adiabatic scenario of perturbation theory, we have the density contrast both in the radiation, baryonic matter and  
dark matter of the cosmic fluid. Radiation in this scale, due to acoustic damping, will diffuse and consequently the baryonic matter results
from strong coupling with radiation will be washed out as well \citep{silk}. The only component from the cosmic fluid that can grow
is the dark matter, which, without any interaction with radiation and baryonic matter, grows under its own gravitational potential. In this section, we emphasize the dark matter component in structure formation. We will discuss the contribution of baryonic matter to the formation of stars and planets in the next section.

The evolution of the density contrast of dark matter for scales larger than Jeans length and smaller than the horizon \citep{martel,paddy} is 
give by 
\begin{equation}
\ddot{\delta} + 2H\dot{\delta} - 4\pi\rho \delta = 0,
\label{dcont}
\end{equation}
where, for simplicity, we do not use the subscript of $"DM"$ for the density contrast of dark matter fluid. Here, the Hubble 
parameter is given by the overall density of matter and radiation and we write it as 
\begin{equation}
\nonumber
\rho_t(x) =\frac{\rho_{eq}}{2}(x^{-3} + x^{-4})
\end{equation}
 where $x= a/a_{eq}$ and $a_{eq}$ is the scale factor at the equality of matter and radiation. For simplicity, we replace time derivatives 
by derivatives with respect to parameter $x$. Then equation (\ref{dcont}) can be written as 
\begin{equation}
x^2H^2\delta^{''}+ \left(2xH^2 + x\frac{\ddot{a}}{a}\right)\delta^{'} - 2\pi\rho_{eq}(x^{-3}+x^{-4})\delta = 0.
\label{dd}
\end{equation}
Also ,we write the FRW equations in terms of $x$:
\begin{eqnarray}
H^2 &=& \frac{4\pi}{3}\rho_{eq}(x^{-3} + x^{-4}) - \frac{k}{a_{eq}^{2}}x^{-2},\\ 
\frac{\ddot{a}}{a} &=& -\frac{2\pi}{3}\rho_{eq}(x^{-3} + 2 x^{-4}).
\end{eqnarray}
By substituting the dynamics of universe in equation (\ref{dd}), the evolution of the density contrast simplifies to 
\begin{equation}
x(1+x + \omega^2 x^2)\delta^{''} + (1+\frac32x+2\omega^2 x^2)\delta^{'}-\frac32(1+\frac{1}{x})\delta = 0.
\label{me}
\end{equation} 
where $\omega^2 = - \frac{3k}{4\pi\rho_{eq}a_{eq}^2}$, and by substituting the definition of spacial curvature we 
rewrite the definition of $\omega$ as 
\begin{equation}
\omega^2 = -(\frac{1}{x}+\frac{1}{x^2})(1-\frac{1}{\Omega}).
\label{omega1}
\end{equation}
We note that $\omega$ is time-invariant and, for the equality time where the density of matter and radiation are equal, $\omega^2 = -2(1-\Omega_{eq}^{-1})$. In the case of $\Omega_{eq}=1$ or $\omega = 0$ in equation (\ref{me}), we 
recover a standard differential equation for the evolution of the density contrast of the dark matter component in a flat universe \citep{mukhanov}. It is also convenient to write equation (\ref{omega1}) in terms of the initial condition of inflation as follows:
\begin{equation}
\omega^2=  (1-\Omega_i) (\frac{a_i}{a_f})^2 (\frac{a_{eq}}{a_f})^2.
\label{eq}
\end{equation}

\begin{figure}
\centering
  \includegraphics[width=85mm]{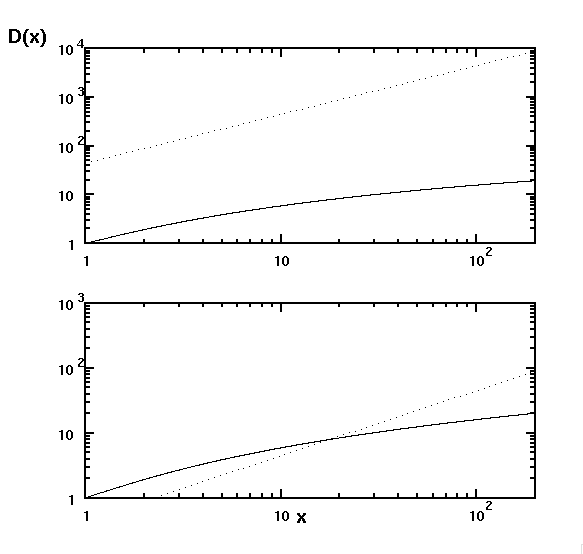}
  \caption{The Growth facto for the density contrast (i.e. $D(x)$) in solid line is compared with the condition for the growth of the structure in (\ref{condition}) (i.e. $\omega^2 x^2/\delta_i(1+x)$) in dotted line. The upper panel is for $\omega = 0.03$ and lower panel is for $\omega = 0.003$. In the lower panel the condition in (\ref{condition}) for a short period of time is fulfilled. For the limit of  $\omega<10^{-4}$ this condition is fulfilled for the structures in the linear regime to grow up to $\delta \simeq 1$ and after that non-linear evolution starts.}
  \label{fig4}
\end{figure}

 Unlike a closed universe, for an open universe structures can grow under specific conditions that depend on the parameters of the open universe. From the spherical top-hat model for structure formation, the energy condition for an overdense patch of the universe that decouples dynamically from the background and 
collapses is $\delta > \Omega^{-1} -1$ \citep{paddy}.  Substituting (\ref{omega1}) in this condition results in 
\begin{equation}
D(x)>\frac{\omega^2 x^2}{\delta_i(1+x)}, 
\label{condition2}
\end{equation}
where $D(x)$ is the growth factor, in which the density contrast relates to the initial value by $\delta(x) = D(x) \delta_i$. In order to investigate this condition, we solve differential equation of (\ref{me}) numerically with the initial conditions at the equality time. The initial perturbation is set to  $\bar\delta_i \simeq 10^{-5}$. Figure (\ref{fig4}) represents the growth factor of density contrast 
(i.e. $D(x)$), which is compared with the condition of collapse of  structures at the right-hand side of (\ref{condition2}). From this investigation, for the critical limit of $\omega<10^{-4}$, this condition is always fulfilled and structures can grow up to the end of linear regime,  where $\delta \simeq 1$, and after that enter the non-linear phase of the structure formation. 

The analytical solution of equation (\ref{me}) for the limit $\omega \rightarrow 0$ is
\begin{eqnarray}
\delta(x) &=& C_1 (\frac23 + x) \\ \nonumber
 &+& C_2(3\sqrt{1+x} - (2+3x)\tanh^{-1}\sqrt{1+x}),
\end{eqnarray}

which confirms linear growth of the structures for $x\gg 1$.  After entering to the non-linear phase, structures decouple from the background expansion of the universe and spherically collapse to non-linear structures; consequently they virialize. The final 
stage of halos, in the case of having enough baryonic budget of the universe, is the formation of stars and planets.

In what follows, we derive the number of e-folding from inflation such that structures can grow in an open universe with the condition of  $\omega<10^{-4}$. In equation (\ref{eq}), we need to know the corresponding scale factor for the equality time. This scale factor 
depends on the ratio of matter to the radiation content of the universe. In other words, this ratio depends on asymmetry between particles and antiparticles. For a given type of particle with the chemical potential $\mu$ and rest mass of $m$, the difference between the density of matter and anti-matter (i.e. $\Delta n$) for a temperature that is smaller than the rest mass of the particle (i.e. $T\ll m$)  \citep{zurek} is
\begin{equation}
\eta = \frac{\Delta n}{n_\gamma} = \frac{n}{n_\gamma} \simeq 1.33\frac{\mu}{T}.
\end{equation}
where $n$ is the net number of matter and $\eta = n/n_\gamma$ for baryons and leptons in the case of our universe is of the order 
of $\sim 10^{-10}$.  Here, for simplicity we define $\eta_{10}$ as $\eta_{10}=\eta/10^{-10}$. In the case of $\eta \rightarrow 0$, where there is complete symmetry between matter and antimatter, the number density of photons does not change considerably. In other words,  $n_\gamma(\eta_{10} =0) \simeq n_\gamma(\eta_{10} =1)$.  In this extreme case, due to absence of matter no structure can form. On the other hand, in the case of strong asymmetry between the matter and anti-matter, the number of dark matter particles as well as baryonic matter particles would be of the same order as that of photons, or $\eta \simeq \mathcal{O}(1)$.

In the case of our universe, where the equality time happens of the order of eV, taking the end of inflation at the GUT scale, the ratio of scale factors at the equality time and the end of inflation would be ${a_{eq}}/{a_f}\sim 10^{23}$ or, in terms of $\omega$,  $\omega^2 \sim 10^{46}(\frac{a_{i}}{a_f})^2$. From the upper bound on $\omega$, e-folding in an open universe where structures can grow has $N_\star>62$, where, from equation (\ref{e}), the probability of this universe is highly probable. 

\begin{figure}
\centering
  \includegraphics[width=80mm]{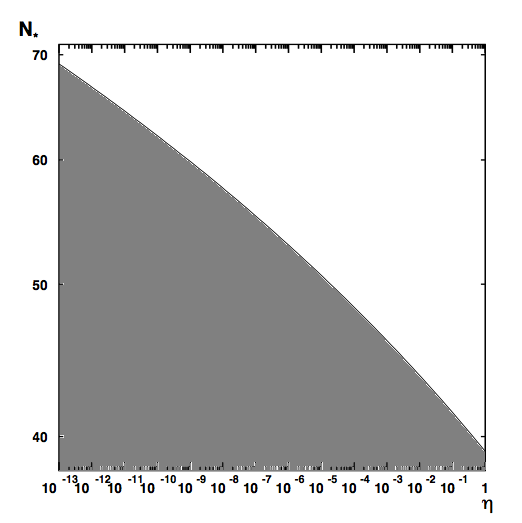}
  \caption{Exclusion area identified by grey shading in terms of $N_\star$ and $\eta$ for the formation of structures based on condition (\ref{condition}). For larger asymmetry between matter and anti-matter (smaller $\eta$), we need a larger $N_\star$ for the formation of structures. }
  \label{fig3}
\end{figure}

In what follows, we determine the e-folding number with an arbitrary value of $\eta$ in the range of $0<\eta<1$ in which structures can grow. Using the general definition of  
$\eta$ from equation (\ref{eta1}), for the equality of matter and radiation, $\rho_X = \rho_\gamma$, $T_{eq} = m_X \eta$. Since the temperature of the universe changes as $1/a$, we can express the scale factor of equality in terms of $\eta_{10}$ as follows:
\begin{equation}
a_{eq}(\eta_{10}) =\frac{a_{eq}(1)}{ \eta_{10}}.
\end{equation}
We substitute this term in equation (\ref{eq}) and, using the condition of $\omega<10^{-4}$, the lower bound on e-folding is obtained as 
\begin{equation}
N_\star> 17\ln 10 - \ln\eta.
\label{condition}
\end{equation}
Figure (\ref{fig3}) represents the domain that satisfies condition (\ref{condition}) for the growth of structures. Smaller asymmetry between matter and anti-matter needs a higher of e-folding to fulfill the structure formation condition.

\section{Baryon asymmetry and selection probability}
\label{ba}

We have seen that the amount of e-folding during inflation has a direct effect on the formation of structures in the universe.  Also, we have investigated the effect of dark matter asymmetry on the formation of structures. In the following section, we investigate the effect of baryonic matter asymmetry in the formation of baryonic structures and consequently life.

 A Standard way to describe the probability of life formation for a given parameter space \citep{tegmark1} is to multiply the independent probability factors. Here we have a three-dimensional parameter space of $N$, $\eta$ (asymmetry in the dark matter) and $\eta_B$ (asymmetry in the baryonic matter) where the probability of a habitable universe can be written as 
 \begin{equation}
 P(N,\eta,\eta_B) = P_{prior}(N,\eta,\eta_B) \times P_{select}(N,\eta,\eta_B), 
 \end{equation}
 where $P_{prior}(N,\eta,\eta_B)$ has defined in section (\ref{inf}), is the theoretical distribution of $N$, $\eta$ and $\eta_B$ at the end of inflation. Since in the inflation there is no explicit dependence between the number of e-folding and the matter asymmetry, we take $N$ as the single parameter that enters the prior probability function. We have seen that the probability of e-folding with $N_\star>60$, which is suitable for the formation of structures (i.e. $P_{prior}(N_\star>60)$) is almost equal to one. On the other hand, $P_{select}(N,\eta_X,\eta_B)$ is proportional to the density of structures such as galaxies and depends on the e-folding, dark matter and baryonic matter content of the universe.

Now, let us assume a universe with enough e-folding during inflation in which almost a flat universe is produced. We impose the selection condition as the sufficient condition for the formation of baryonic structures. What are the bounds on baryon asymmetry of the universe to form stars and planets ? Here, the free parameter of our concern is the ratio of the number density of baryons to the number density of photons, $\eta_B= n_B/n_\gamma$.

For an adiabatic perturbation mode that enters the horizon, baryonic matter is mainly under the influence of the dark matter potential and  evolution of density contrast for baryonic matter is given by 
\begin{equation}
a^2\delta_B^{''} +\frac32 a\delta_B^{'} =\frac32 \delta_D, 
\end{equation}
  where the prime denoted a derivative with respect to the scale factor. This equation has the solution $\delta_B = \delta_{D}^{(en)}a/a_{en}$ where "en" corresponds to the entry time of a perturbation mode into the horizon. Then, at the time at which a dark matter halo becomes non-linear (i.e. $\delta_D\sim 1$), the density contrast of the baryonic matter also approaches to one.
  
 In the non-linear phase of structure formation, assuming a top-hat spherical model, each shell collapses without crossing the other shells and at turnaround time the structure start collapsing and finally virialize \citep{paddy}. If we ignore the dissipation of the baryonic matter at the first stage of virialization, then baryonic matter particles will have almost the same amount of kinetic energy and potential energy as the 
 dark matter particles and will virialize at the same time with the dark matter fluid. Then, in the gravitational potential of dark matter, we expect that the baryonic matter extends to the same scale as the dark matter structure. Now we can calculate the density of baryonic matter in the dark matter halo from the following relation: 
 \begin{equation}
 \rho_B = \rho_D\frac{m_p}{h\nu_0}\frac{\Omega_R^{0}}{\Omega_M^{0}}\eta_B, 
\label{rho}
 \end{equation}
 where $h\nu_0$ is the energy of CMB photons, $m_p$ is the mass of proton and "zero" sup/superscript corresponds to the value of parameters at the present time.

 The important issue concerning the amount of baryonic matter within the gravitational potential of dark matter is the possibility 
 of formation of a baryonic structure. Formation of baryonic structure after the virialization of halo is possible if plasma can cool and 
 fall to the centre of the gravitational potential of the dark matter. The essential criterion is comparing the dynamical time-scale with the cooling time-scale of the baryonic matter. The dynamical time-scale for a test particle in the potential of the dark matter is given by
 \begin{equation}
 t_{grav} = (\frac{R^3}{GM_D})^{1/2},
 \end{equation}
and, on the other hand, the cooling time-scale is 
 \begin{equation}
 t_{cool} = (n_e \alpha \sigma_T)^{-1}(\frac{T}{m_e})^{1/2},
 \end{equation}
 where $\sigma_T$ is the Thomson cross-section and $\alpha$ is the fine-structure constant. For $t_{cool}<t_{grav}$, we have a sufficient condition for the formation of baryonic structures. Substituting 
 temperature from the virial condition, we get the simple result
 \begin{equation}
 M_B = \frac{R}{74~kpc} M_D,
 \label{rho2}
 \end{equation}
 where baryonic matter and dark matter are enclosed within the same radius $R$. Integrating from equation (\ref{rho}) and equating it with equation (\ref{rho2}), we obtain the following relation between 
 the size of structure and the mass of baryonic matter in terms of $\eta_B$:
 \begin{eqnarray}
 R &=& 8.15 \eta_{10,B}~\text{kpc}, \\ 
 M_B &=& 0.11 \eta_{10,B} M_D
 \label{mb}
 \end{eqnarray}
 where $\eta_{10,B}$ is the definition of $\eta_{B}$ normalized to $10^{-10}$.  For a Galactic halo with the mass of the 
 order of $M_D \simeq 10^{12} M_\odot$, for $\eta_{10,B}>10^{-5}$, a structure with the mass of a globular cluster can form within this 
 dark halo.  Another lower value for the baryonic asymmetry could be $\eta_{10,B}\simeq 10^{-12}$, where only one star with $M_{min} \sim 0.1 M_\odot$ condenses to form a single star within a $10^{12}$ solar mass halo. We can think about another extreme case, having only one star within the cosmological horizon. In this case, we can write the $\eta_B$ parameter in terms of minimum mass for a star and the mass of matter within the horizon (i.e. $M_H$), 
 \begin{equation}
 \eta_{10,B} = \frac{m_X}{m_p}\frac{M_{min}}{M_H} \eta_{10},
 \label{silk}
 \end{equation}
  where $m_X$ and $m_p$ are the mass of dark matter particle and proton mass. Substituting the numerical values for these parameters, we obtain $\eta_{10,B} = 10^{-24}$.  At this extreme small value for $\eta_{10,B}$, we  will have only one star in the observable universe.

   A stronger astrophysical constrain has been studied, using the amount of minimum baryonic matter for formation of a baryonic disc in the galaxies  \citep{tegmark1}. We note that the upper bound of the baryon contribution to the matter content of the universe is also limited by Silk damping for $f_B = M_B/M_D<1/2$ \citep{tegmark1} where, in terms of $\eta_{B}$, we can write 
 \begin{equation} 
 \frac{\eta_B}{\eta}\frac{m_p}{m_X}<1/2.
 \label{bb}
 \end{equation}
 Assuming the mass of dark matter particles is of the order of $10$ GeV and $\eta\simeq 10^{-10}$ for the dark matter component, the amount of asymmetry in baryonic matter from Silk damping would be $\eta_{B}<10^{-9}$.
 
We can also derive another constrain from nearby explosions of supernovas, which can cause mass extinction on an Earth-like planet. A detailed study by \cite{ellis} shows that supernovas at distances shorter than $10 pc$ could produce a flux of energetic electromagnetic and charged particles that destroys Earth's ozone layer over a period of $300$~yr. Assuming that the rate of supernova explosions for a galaxy with $10^{11}$ stars is about one explosion per $10$ years, we can write the rate of supernova explosion as 
\begin{equation}
\Gamma_{exp} =  10^{-12} star^{-1}~ yr^{-1}. 
\end{equation}
The typical density of stars in the Milky Way is about $n\simeq 1$ star$/pc^{-3}$. For a given baryonic asymmetry, the number density of stars will scale with this factor as $n\simeq \eta_{10,B} $ star$/pc^{-3}$. The rate of explosions within the sphere with the radius smaller than $10~pc$ would therefore be 
\begin{equation}
\frac{dN(<10pc)}{dt} = 4\eta_{10,B}\times 10^{-9}~yr^{-1}.
\label{rate}
\end{equation}
In the case of $\eta_{10,B} =1$, \cite{ellis} estimate that, during the Phanerozoic era with a duration of $\Delta t=240$~Myr, we would expect one explosion than may cause mass extinction on Earth. 

For the extreme case where the ozone layer undergoes bombardment of high-energy particles with a sequence of explosions, we assume the rate of one explosion per $300$ years. In this case, life never emerges on a planet. Substituting this rate in the left hand side of equation (\ref{rate}), the maximum value for the baryonic asymmetry obtained is $\eta_B = 10^{-4}$, so from the supernova explosions we put an upper bound of $\eta_B<10^{-4}$ on this parameter.

\begin{figure}
\centering
  \includegraphics[width=80mm]{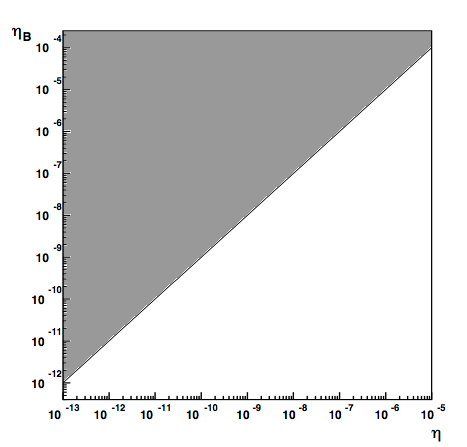}
  \caption{Excluding area in $\eta$, $\eta_B$ space from the Silk damping constrain in equation (\ref{bb}). We also exclude $\eta_B>10^{-4}$ from the effect of supernova explosions on life.}
  \label{ee}
\end{figure}

From this section, we can conclude that a larger $\eta_{B}$ produces more baryonic structures; however, it has been limited to $\eta_B <10 \eta$ from equation (\ref{bb}) for the Silk damping limit. Moreover, $\eta_B$ is limited by the effect of frequent supernova explosions not being larger than $10^{-4}$. Figure (\ref{ee}) shows the exclusion area in $\eta_B,\eta$ space. Since the number of stars and planets is proportional to $\eta_B$, the probability of selection depends roughly in a linear way on $\eta_B$, with the cut-off at $\eta_B<10^{-4}$, 
 \begin{equation}
 P_{select}(\eta) \propto \eta_B~~~~~ ,\eta_B<10^{-4}
 \end{equation} 
Having a fixed value of $\eta$ for dark matter, we would obtain a given number density of halos 
from the Press-Schechter (PS) function in a flat universe  \citep{ps}; however, the baryonic matter associated with the dark halos would be scaled by $\eta_B$.

 \section{conclusion}
 \label{conc}

In this work, from the initial condition point of view, we studied the probability of a universe being habitable. We have split this probability into prior and selected probabilities. The prior probability depends on the number of e-folding during inflation. We have shown that, for a power-law inflaton field, the probability for e-folding of the order of $60$ is close to one, which means that the formation of universes similar to ours is highly probable. 

Next, assuming an open or closed universe, we investigated the growth of structures in such universes. For the case of a closed universe, due to the negative energy condition of the overdense regions, all structures are able to grow until they collapse to a virialized regime. We note that the limiting parameter in this case is the age of the universe. We could constrain the e-folding to be larger than $\sim 60$ for the formation of life. In an open universe, we solved the growth of density contrast numerically and compared it with the energy condition that allows for the non-linear phase for the growth of structures. In an open universe, we obtained almost the same e-folding number for the formation of life as in the closed universe.

Finally we have investigated the asymmetry parameter between baryons and 
anti-baryons. In order to form a single structure of the order of a globular cluster within the Milky Way halo, the asymmetry parameter needs to be $\eta_{10,B}>10^{-5}$ and, for the extreme case where we have only one star within the Hubble radius, we obtain the minimum bound of $\eta_{10,B}>10^{-24}$.  We note that the physical condition for the formation of disc galaxies and stars is much higher than this limit. Also, the upper bound for the fraction of baryons is limited by Silk damping with the constraint $\eta_B<10\eta $. We have also investigated the constraint arising from nearby explosions by supernovae, which can destroy life in nearby planets. Increasing $\eta_B$ increases the number density of stars in the galaxy and, from the rate of supernova explosions and a limiting distance of $10$ pc away from the supernova, we obtained an upper limit on the amount of baryonic matter asymmetry as $\eta_B<10^{-4}$.

Concluding this work, the overall probability of a universe being habitable is a produce of the prior and selection probabilities, where the prior probability function is almost unity from inflation and the selection part is proportional to $\eta_B$. In another words, the probability of a universe being habitable is proportional to the baryonic asymmetry of the universe (i.e. $P\propto \eta_B$), with a cutoff at $\eta_B<10^{-4}$.

%In this part we have shown that the probability of baryonic structure formation which is proportional to formation of life is proportional to the asymmetry parameters of baryons (i.e. $P_{selected} \propto \eta_B$). 

%necessary and sufficient condition for the formation of structures in the universe by means of formation of stars and consequently the planets. We have shown 
%that necessary condition for the growth of the structures with an arbitrary pre-inflation initial condition is that e-folding during 
%the inflation should be at least larger than $57$. 

 %The sufficient condition is that the asymmetry between the 

\section*{Acknowledgements}

I thank the anonymous referee for useful comments and guides to 
on how to improve this work. This work was supported by Sharif University of Technology's office of Vice President for Research under Grant No. G950214.

\label{lastpage}

\end{document}